\documentclass[apjl]{emulateapj}

\begin{document}

\shorttitle{DoAr 25 Disk Structure}

\shortauthors{Andrews et al.}

\title{The Structure of the D\lowercase{o}A\lowercase{r} 25 Circumstellar Disk}

\author{Sean M. Andrews\altaffilmark{1}, A. M. Hughes, D. J. Wilner, and 
Chunhua Qi}

\affil{Harvard-Smithsonian Center for Astrophysics, 60 Garden Street, 
Cambridge, MA 02138}
\altaffiltext{1}{Hubble Fellow}

\begin{abstract}
We present high spatial resolution ($\lesssim 0\farcs3 \approx 40$\,AU) 
Submillimeter Array observations of the 865\,$\mu$m continuum emission from the 
circumstellar disk around the young star DoAr 25.  Despite its bright 
millimeter emission, this source exhibits only a comparatively small infrared 
excess and low accretion rate, suggesting that the material and structural 
properties of the inner disk may be in an advanced state of evolution.  A 
simple model of the physical conditions in the disk is derived from the 
submillimeter visibilities and the complete spectral energy distribution using 
a Monte Carlo radiative transfer code.  For the standard assumption of a 
homogeneous grain size distribution at all disk radii, the results indicate a 
shallow surface density profile, $\Sigma \propto r^{-p}$ with $p \approx 0.34$, 
significantly less steep than a steady-state accretion disk ($p = 1$) or the 
often adopted minimum mass solar nebula ($p = 1.5$).  Even though the total 
mass of material is large ($M_d \approx 0.10$\,M$_{\odot}$), the densities 
inferred in the inner disk for such a model may be too low to facilitate any 
mode of planet formation.  However, alternative models with steeper density 
gradients ($p \approx 1$) can explain the observations equally well if 
substantial grain growth in the planet formation region ($r \le 40$\,AU) has 
occurred.  We discuss these data in the context of such models with dust 
properties that vary with radius and highlight their implications for 
understanding disk evolution and the early stages of planet formation.  
\end{abstract}

%\keywords{circumstellar matter --- planetary systems: protoplanetary disks --- 
%stars: formation --- stars: pre-main$-$sequence --- stars: individual (DoAr 25)}

\section{Introduction}

The details of the planet formation process in its early stages are largely 
determined by the physical conditions in the disk around the host star.  The 
densities, temperatures, material content, and structure of this gas and dust 
reservoir critically influence the mechanisms and efficiencies of the assembly 
of a planetesimal population, as well as its subsequent growth and dynamical 
evolution.  Spatially resolved observations of disks at millimeter wavelengths 
provide unique access to these physical conditions, including in particular the 
spatial distribution of mass \citep[e.g.,][]{wilner00}.  Such constraints on 
the physical structure of disks around young ($\sim$1\,Myr) T Tauri stars are 
expected to offer a glimpse at the relatively pristine {\it initial} conditions 
available for planet formation \citep{beckwith96}.

However, recent observations of some disks raise questions about how pristine 
or initial these conditions really are, even at such young ages.  A small 
collection of ``transition" disks show pronounced dips in their infrared 
spectral energy distributions (SED) which suggest that large inner holes or 
gaps have been cleared \citep[e.g.,][]{calvet05,espaillat07}; a scenario 
confirmed in some cases by direct imaging \citep{pietu06,hughes07,brown08}.  
Although less dramatic, a separate population of disks with relatively faint 
infrared SED excesses (although without distinct SED gaps) but bright 
millimeter emission can be interpreted as evidence for substantial particle 
growth $-$ and thus diminished grain emissivity $-$ in the inner disk 
\citep{lada06,najita07}.  These indicators suggest that some fraction of 
$\sim$1\,Myr-old disks have already made significant progress along the path 
toward making planets.

In this Letter, we use high spatial resolution submillimeter observations to 
help interpret the physical conditions in the disk around DoAr 25.  Located in 
the L1688 dark cloud in Ophiuchus \citep[$d \approx 145$\,pc;][]{makarov07}, 
the DoAr 25 disk has a comparatively small infrared excess and accretion rate 
\citep[\emph{\.{M}} $\sim$10$^{-10}$-10$^{-9}$\,M$_{\odot}$ 
yr$^{-1}$;][]{muzerolle98,natta06}, but bright millimeter emission 
\citep{aw07b}.  Following a description of the observations and their 
calibration in \S 2, we utilize a standard disk model and radiative transfer 
calculations to estimate the disk structure and mass distribution in \S 3.  We 
discuss the results in the contexts of particle growth and planet formation in 
\S 4.

\section{Observations and Data Reduction}

\begin{figure*}
\epsscale{1.0}
\plotone{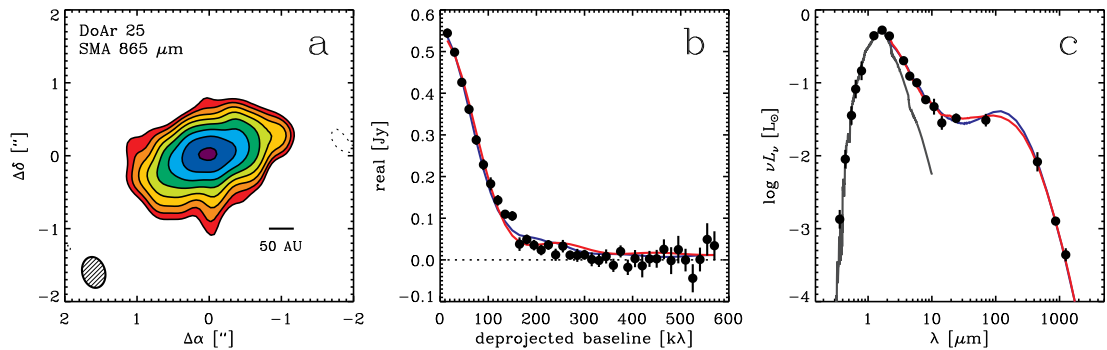}
\caption{({\it left}) High resolution SMA image of the 865\,$\mu$m continuum
emission from the DoAr 25 disk.  Contours start at 7.5\,mJy beam$^{-1}$
(3\,$\sigma$) and increase by factors of $\sqrt{2}$.  The synthesized beam size
is shown in the lower left corner.  ({\it middle}) The elliptically-averaged
865\,$\mu$m visibilities (the real part) as a function of the deprojected
baseline length.  ({\it right}) The DoAr 25 SED, where the ordinate corresponds
to $4 \pi d^2 \nu F_{\nu}$.  Optical photometry is from \citet[][$UBV$]{vrba93}
and \citet[][$RI$]{wilking05}.  Infrared data are from 2MASS \citep{cutri03},
the {\it Spitzer Space Telescope} (L.~E.~Allen 2007, {\it private
communication}), \citet{barsony05}, and \citet{bontemps01}.  Millimeter fluxes
are from \citet{dent98}, this paper, and \citet{andre94}.  Error bars include
uncertainties in the absolute flux scales.  The grey curve shows the Kurucz
model adopted as the input stellar radiation source for the radiative transfer
modeling.  Overlaid on the visibility and SED panels are the model disk
behaviors for a standard grain size distribution ({\it red}: \S 3) and an
alternative distribution ({\it blue}) with grain growth (and reduced
emissivity) in the inner disk ($r < 40$\,AU; \S 4).\label{SMA_image}}
\end{figure*}

DoAr 25 ($\alpha = 16^{\mathrm h}26^{\mathrm m}23\fs68$, $\delta = 
-24\degr43\arcmin14\farcs1$ [J2000]) was observed in the ``very extended" 
configuration of the Submillimeter Array \citep[SMA;][]{ho04} on 2007 May 26 
and June 17 in exceptional conditions, with zenith optical depths of $<$0.06 at 
225\,GHz.  In this arrangement, the 8 SMA antennas (6\,m diameter each) span 
physical baselines up to 508\,m.  The observations cycled between DoAr 25 and 
two quasars (J1517$-$243 and J1626$-$298) on 10 minute intervals.  The raw 
visibilities were calibrated with the MIR package.  Passband calibration was 
conducted with bright quasars (3C273, 3C279), and complex gain calibration was 
performed with J1626$-$298.  MWC 349 and Callisto were used to set the absolute 
flux scale, which is accurate to $\sim$10\%.  Data from both nights and each of 
the two 2\,GHz sidebands were combined.  Using J1517$-$243 as a consistency 
check on the phase transfer, we estimate that phase noise and baseline errors 
generate an effective seeing of $\lesssim$0\farcs1. 

Additional observations in the compact and extended SMA configurations 
(baselines of $\sim$15-200\,m) were obtained in 2006 \citep[see][]{aw07}.  The 
data show excellent consistency between all configurations on overlapping 
spatial scales.  All continuum data were combined to provide visibilities at an 
effective frequency of 346.9\,GHz (865\,$\mu$m).  For the 2006 data, we 
simultaneously observed the CO $J$=3$-$2 transition at 345.796\,GHz with a 
spectral resolution of 0.70\,km s$^{-1}$.  The MIRIAD package was utilized for 
the standard tasks of Fourier inversion, deconvolution, and imaging of the 
calibrated visibilities.  A submillimeter continuum map was generated with a 
Briggs robust=0.25 weighting scheme, providing a synthesized beam FWHM of 
$0\farcs43\times0\farcs32$ at a position angle (PA) of 15\degr.

\section{Results}

Figure \ref{SMA_image}a shows a high spatial resolution 865\,$\mu$m continuum 
image of the DoAr 25 disk.  An elliptical Gaussian fit to the visibilities 
indicates a well-resolved source with an integrated flux density of 
$563\pm3$\,mJy (not including the calibration uncertainty) and an inclination 
of $62\pm3\degr$ at a PA of $111\pm3\degr$.  The SMA visibilities are presented 
in Figure \ref{SMA_image}b, elliptically averaged after their deprojection 
based on the above geometry \citep[see][]{lay97}.  The visibilities drop off 
steeply, suggesting a shallow gradient in the radial surface brightness profile 
of the disk.  The SED for DoAr 25 is shown in Figure \ref{SMA_image}c, using 
data compiled from the literature.  A relatively small thermal excess is noted 
above the stellar photosphere out to mid-infrared wavelengths, before the SED 
picks up to show some of the brightest millimeter emission of all the T Tauri 
disks in Ophiuchus \citep{aw07b}.

For a quantitative view of the physical conditions in the DoAr 25 disk, we 
attempted to reproduce the full SED and SMA visibilities with the 2-D 
axisymmetric Monte Carlo radiative transfer code RADMC (v3.1, C.~P.~Dullemond 
2007, {\it private communication}).  This code utilizes the algorithm developed 
by \citet{bjorkman01} to compute temperatures in a circumstellar medium heated 
solely by stellar irradiation and with a density structure 
\begin{equation}
\rho (r,\theta) = \frac{\Sigma}{\sqrt{2\pi} H} \exp{\left[ -\left( \frac{r \cos{\, \theta}}{\sqrt{2}H} \right)^2 \right]},
\end{equation}
where the radial coordinate ($r$) runs from the dust sublimation radius to an 
outer boundary ($R_d$), the angular coordinate ($\theta$) runs from the 
rotation axis ($\theta = 0$) to the midplane ($\theta = \pi/2$), and the 
surface density and scale height vary with radius as power laws ($\Sigma = 
\Sigma_0 (r/R_d)^{-p}$ and $H = H_0 (r/R_d)^h$).  Aside from some minor 
coordinate conventions, this disk structure model is identical to others 
commonly used in the literature \citep[e.g.,][]{chiang97}.  The code itself was 
described by \citet{dullemond04}, but now includes a diffusion algorithm that 
is employed for grid locations with poor photon statistics (i.e., near the disk 
midplane).  

The central illumination source was a Kurucz model (gray curve in Fig.~1c) 
fixed to match the properties of the star and optical SED: a K5 spectral type 
($T_{\rm{eff}}$=4250\,K), $L_{\ast}$=1.3\,L$_{\odot}$, and $A_V$=2.9 
\citep{wilking05}.  We initially adopted the emissivity properties provided by 
\citet{dalessio01}, for a dust grain size ($a$) distribution that varies as 
$n(a) \propto a^{-3.5}$ between $a_{\rm{min}}$=0.005\,$\mu$m and 
$a_{\rm{max}}$=1\,mm.  These emissivities are essentially identical to the 
standard disk values used at millimeter wavelengths \citep{beckwith90}.  With 
this information, we ran the radiative transfer code over a grid of 5 free 
parameters that characterize the disk structure: $R_d$, $p$, $h$, and the 
normalizations of $\Sigma$ (i.e., the disk mass, $M_d$) and $H$.  For each 
point in the grid a model SED and 865\,$\mu$m visibilities were generated with 
a raytracing program, assuming the above orientation.

\begin{figure}
\epsscale{1.15}
\plotone{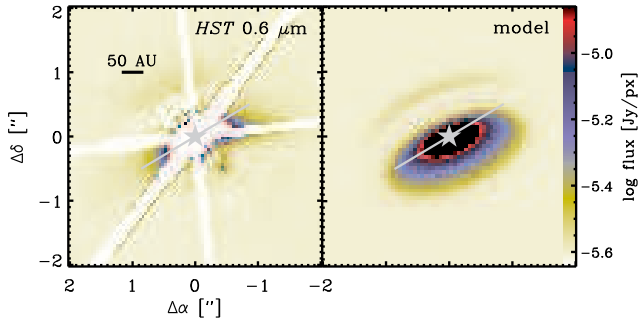}
\caption{({\it left}) An optical {\it HST} image (F606W, $\sim$0.6\,$\mu$m) of
starlight scattered off the surface of the DoAr 25 disk.  ({\it right}) The
scattered light image predicted from the best-fit model parameters described in
\S 3, based solely on the SMA visibilities and the SED.  The stellar position
is labeled with a star symbol, a gray line marks the orientation of the disk
major axis, and a logarithmic brightness scale is shown on the right.  While
the model does not include direction-dependent scattering (and is therefore not
used in the structure determination), the comparison with the {\it HST} data
confirms the size scales, orientation, and general disk geometry that were
derived from the SMA data alone.
\label{scatlight}}
\vspace{0.1cm}
\end{figure}

The best-fit model parameter values estimated from minimizing $\chi^2$ are $R_d 
= 310$\,AU, $p = 0.34$, $h = 1.11$, $M_d = 0.10$\,M$_{\odot}$, and $H = 28$\,AU 
(at $R_d$).  The synthetic SED and visibilities for a model with these 
parameters are shown in red in Figure 1.  The reduced $\chi^2$ value comparing 
the best-fit model and the data shown in Fig.~1b and 1c was $\sim$2, primarily 
due to difficulty reproducing the mid-infrared spectrum in detail.  The formal 
1\,$\sigma$ parameter uncertainties are $\sim$15\,AU for $R_d$, 
$\sim$0.01\,M$_{\odot}$ for $M_d$, and $\sim$5\,AU for $H$ (at $R_d$), while 
$p$ can range from 0.2$-$0.4 and the flared shape of the disk can run as low as 
$h \approx 1.07$ (but not higher than the best-fit value without over-producing 
infrared emission).

As an {\it a posteriori} test of the derived disk properties, Figure 
\ref{scatlight} compares an optical (0.6\,$\mu$m) {\it Hubble Space Telescope} 
({\it HST}; WFPC2/F606W) image of starlight scattered off the DoAr 25 disk 
surface \citep{stapelfeldt08} with a RADMC model prediction corresponding to 
the above parameters.  Because this version of RADMC only treats isotropic 
scattering, a quantitative comparison with the data is unjustified.  However, 
it is clear that the best-fit model is in good qualitative agreement with the 
disk orientation and geometry as traced by both the scattered flux scale and 
the location and intensity of the dust lane marking the midplane.

While not shown here, channel maps of the CO $J$=3$-$2 transition from the SMA 
data (with a $1\farcs8 \times 1\farcs6$ beam at PA = 4\degr) exhibit two 
distinct patches of weak line emission separated by $\sim$4\,km s$^{-1}$ toward 
DoAr 25, centered on the local cloud velocity ($V_{{\rm LSR}} = 3.3$\,km 
s$^{-1}$).  Single-dish mapping of the Ophiuchus clouds in the CO $J$=1$-$0 
transition shows a bright, optically thick line with a FWHM of $\sim$3\,km 
s$^{-1}$ coincident with the DoAr 25 position \citep{ridge06}.  Unfortunately, 
this suggests that the line emission from the disk remains substantially 
contaminated by confusion with the more extended molecular cloud.

\section{Discussion}

The spatial distribution of mass, characterized by the surface density index 
$p$, is a critical diagnostic for understanding the material evolution and 
planet formation potential of a disk.  It reflects the disk viscosity and 
therefore provides an observable indication of the evolution timescale dictated 
by the accretion process and angular momentum conservation.  Moreover, the mass 
distribution will influence the composition and architecture of a developing 
planetary system.  By comparing the decay of accretion rates over time with 
models that use a simple prescription for the turbulent viscosity, 
\citet{hartmann98} argue for a typical value $p$=1.  A steeper gradient 
($p$=1.5) is invoked to account for the mass distribution in the solar system, 
once augmented to cosmic abundances and smeared into annuli \citep[the minimum 
mass solar nebula, or MMSN;][]{weidenschilling77}.

\begin{figure}
\epsscale{0.92}
\plotone{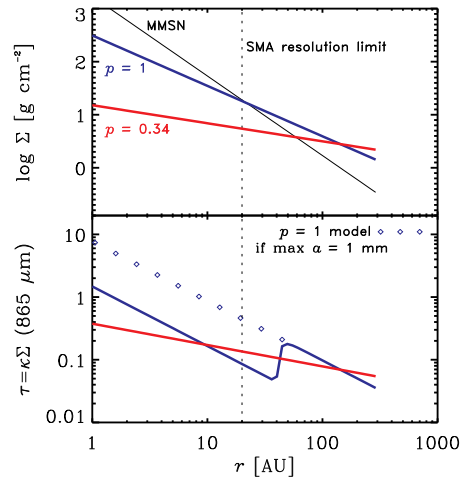}
\caption{({\it top}) The radial surface density profile ($\Sigma$; gas+dust)
for the best-fit model described in \S 3 ({\it red}) and the toy model with
large grains in the inner disk described in \S 4 ({\it blue}).  The latter is
decomposed into two grain populations, with $a_{{\rm max}} = 1$\,mm outside
40\,AU and 1\,m \citep[both from][]{dalessio01} inside that radius.  The
minimum mass solar nebula (MMSN) density profile is shown in black for
comparison.  ({\it bottom}) The radial profile of the 865\,$\mu$m continuum
optical depth ($\tau = \kappa \Sigma$; where $\kappa$ represents the material
emissivity) for the same models.  The open diamonds track the optical depth
that would be expected if the standard ($a_{{\rm max}} = 1$\,mm) grain size
distribution emissivity were adopted everywhere in the disk for the steep
density gradient ($p = 1$) model.  The SMA resolution limit for these data is
marked with a vertical dotted line.  \label{model_info}}
\end{figure}

The low value of $p$ derived for the DoAr 25 disk would have important 
implications for understanding its evolution.  In the \citet{hartmann98} 
prescription for a viscous accretion disk, such a low value of $p$ would 
produce a significantly diminished mass accretion rate onto the star over most 
of the disk lifetime, perhaps lower by a factor of $\sim$2 compared to the 
canonical disk with $p = 1$.  This is in reasonable agreement with constraints 
on \emph{\.{M}}, which vary from low \citep[$5\times10^{-9}$\,M$_{\odot}$ 
yr$^{-1}$;][]{muzerolle98} to negligible values 
\citep[$< 5\times10^{-10}$\,M$_{\odot}$ yr$^{-1}$;][]{natta06}.  In essence, 
the accretion process is less efficient in a disk with a shallow density 
gradient.  The same could be said about the formation of planets in such a 
disk.  In spite of the large $M_d$, the inner disk densities implied by this 
low-$p$ model are more than an order of magnitude smaller than those required 
by planet formation models \citep[e.g.,][also note that even though 
$M_d/M_{\ast} \approx 0.18$, the Toomre stability criterion holds at all radii 
in this model]{hubickyj05}.  Previous millimeter observations of disks have 
generally found large $p$ \citep[$\ge 0.7$; e.g.,][]{wilner96,aw07,pietu06,hamidouche07}, although the derived value for DoAr 25 lies in the low end 
of a wide distribution; the variety of modeling techniques and spatial 
resolutions in those studies make direct comparisons difficult.

Of course, there are alternative interpretations of the observed DoAr 25 disk 
properties that do not call for such a small density gradient.  The absence of 
a pronounced dip feature in the infrared SED or an apparent ``hole" in the 
submillimeter surface brightness distribution can effectively rule out a 
substantially cleared inner region like those those in transition disks.  
However, because the millimeter continuum emission traces the optical depth in 
the disk $-$ the {\it product} of the density structure and the dust emissivity 
$-$ the observed shallow brightness profile could be explained equally well 
with steep density gradients ($p \ge 1$) if the emissivity compensates by 
decreasing toward small radii.  For the models described in \S 3 (and generally 
in the literature), the grain size distribution is assumed to be spatially 
uniform throughout the disk.  In reality, particle sizes in the inner disk 
might be larger than in the outer disk because the collisional growth timescale 
is shorter for higher densities and orbital velocities \citep[e.g.,][and 
references therein]{dullemond05,garaud07}.  \citet{dalessio01} show that the 
millimeter emissivity quickly decreases as $a_{{\rm max}}$ grows beyond 
$\sim$1\,mm.  Therefore, a shorter grain growth timescale in the inner disk 
would naturally produce an emissivity that decreases at smaller radii.  

To illustrate the effects this would have observationally, we constructed a 
crude toy model that has the standard steady-state accretion disk density 
profile ($p = 1$), but allows particles with a factor of $\sim$5 smaller 
emissivity (at 865\,$\mu$m; corresponding to $a_{{\rm max}} \approx 1$\,m) to 
account for essentially all of the mass on solar system scales ($r \le 
40$\,AU).  As shown in blue in Figure \ref{SMA_image}, models that mimic grain 
growth can reproduce both the DoAr 25 SED behavior and the shallow millimeter 
brightness profile fairly well.  However, given the additional freedom that 
comes with more model parameters, the data do not uniquely distinguish values 
of inputs like $a_{{\rm max}}$.  Different values can be accomodated by 
adjusting the relative abundances and locations of the grain populations.  
Figure \ref{model_info} shows the radial distributions of $\Sigma$ and the 
865\,$\mu$m optical depth for both this toy model and the more standard 
prescription adopted in \S 3.  While the details in such an {\it ad hoc} 
demonstration model are unimportant, the lower panel of Figure \ref{model_info} 
highlights the salient point: a decrease in the inner disk emissivity can 
effectively mask a steep density gradient.

There is considerable evidence from the shape of millimeter continuum spectra 
that disk emissivities have been modified by substantial particle growth, 
albeit in an average sense due to limited spatial resolution 
\citep[e.g.,][]{beckwith91,testi01,testi03,wilner05,rodmann06}.  DoAr 25 has a 
fairly typical long-wavelength SED shape ($n = 2.7\pm0.2$, where $F_{\nu} 
\propto \nu^n$) for a T Tauri star.  High spatial resolution observations at 
multiple frequencies (from $\sim$40-400\,GHz) can potentially disentangle the 
millimeter SED shape ($n$; i.e., colors) in the evolved inner disk from the 
more primordial material at larger radii.  Such data could be used to 
empirically reconstruct the shape of the emissivity spectrum as a function of 
location in the disk, which could subsequently be modeled to map out the 
constituent grain populations.

If the properties of disks like DoAr 25 are caused by particle growth in the 
inner disk, they can be considered less-evolved versions of ``transition" 
disks, with larger remnant optical depths in their inner holes (effectively 
muting out the signature dip in the SED noted for transition disks).  It 
would be interesting to model high-resolution millimeter observations of 
similar disks and see if shallow density gradients (low $p$) are commonly 
inferred when assuming a standard grain size distribution.  If that is the 
case, such results could be a signature that a few percent of $\sim$1\,Myr-old 
disks are already well on their way to making planets.

\acknowledgments We are grateful to Lori Allen, Karl Stapelfeldt, John Krist, 
Melissa McClure, and Dan Watson for kindly sharing their data prior to 
publication, Paola D'Alessio for providing dust emissivity tables, and 
especially to Kees Dullemond and Roy van Boekel for all of their prompt and 
patient assistance with the RADMC code.  The SMA is a joint project between the 
Smithsonian Astrophysical Observatory and the Academia Sinica Institute of 
Astronomy and Astrophysics and is funded by the Smithsonian Institution and the 
Academia Sinica.  Support for this work was provided by NASA through Hubble 
Fellowship grant \#HF-01203.01-A awarded by the Space Telescope Science 
Institute, which is operated by the Association of Universities for Research in 
Astronomy, Inc., for NASA, under contract NAS 5-26555.


\begin{thebibliography}{}
\bibitem[Andr{\'{e}} \& Montmerle(1994)]{andre94} Andr{\'{e}}, P., \& Montmerle, T. 1994, \apj, 420, 837
\bibitem[Andrews \& Williams(2007a)]{aw07} Andrews, S. M., \& Williams, J. P. 2007, \apj, 659, 705 (2007a)
\bibitem[Andrews \& Williams(2007b)]{aw07b} --------- 2007, \apj, 671, 1800 (2007b)
\bibitem[Barsony et al.(2005)]{barsony05} Barsony, M., Ressler, M. E., \& Marsh, K. A. 2005, \apj, 630, 381
\bibitem[Beckwith et al.(1990)]{beckwith90} Beckwith, S. V. W., Sargent, A. I., Chini, R., \& G{\"{u}}sten, R. 1990, \aj, 99, 924
\bibitem[Beckwith \& Sargent(1991)]{beckwith91} Beckwith, S. V. W., \& Sargent, A. I. 1991, \apj, 381, 250
\bibitem[Beckwith \& Sargent(1996)]{beckwith96} --------- 1996, Nature, 383, 139
\bibitem[Bjorkman \& Wood(2001)]{bjorkman01} Bjorkman, J. E., \& Wood, K. 2001, \apj, 554, 615
\bibitem[Bontemps et al.(2001)]{bontemps01} Bontemps, S., et al. 2001, \aap, 372, 173
\bibitem[Brown et al.(2008)]{brown08} Brown, J. M., Blake, G. A., Qi, C., Dullemond, C. P., \& Wilner, D. J. 2008, \apjl, in press
\bibitem[Calvet et al.(2005)]{calvet05} Calvet, N., et al. 2005, \apj, 630, L185
\bibitem[Chiang \& Goldreich(1997)]{chiang97} Chiang, E. I., \& Goldreich, P. 1997, \apj, 490, 368
\bibitem[Cutri et al.(2003)]{cutri03} Cutri, R. M., et al. 2003, 2MASS All-Sky Point Source Catalog (Pasadena: IPAC)
\bibitem[D'Alessio et al.(2001)]{dalessio01} D'Alessio, P., Calvet, N., \& Hartmann, L. 2001, \apj, 553, 321
\bibitem[Dent et al.(1998)]{dent98} Dent, W. R. F., Matthews, H. E., \& Ward-Thompson, D. 1998, \mnras, 301, 1049
\bibitem[Dullemond \& Dominik(2004)]{dullemond04} Dullemond, C. P., \& Dominik, C. 2004, \aap, 417, 159
\bibitem[Dullemond \& Dominik(2005)]{dullemond05} --------- 2005, \aap, 434, 971
\bibitem[Espaillat et al.(2007)]{espaillat07} Espaillat, C., et al. 2007, \apj, 670, L135
\bibitem[Garaud(2007)]{garaud07} Garaud, P. 2007, \apj, 671, 2091
\bibitem[Hamidouche et al.(2006)]{hamidouche07} Hamidouche, M., Looney, L. W., \& Mundy, L. G. 2006, \apj, 651, 321
\bibitem[Hartmann et al.(1998)]{hartmann98} Hartmann, L., Calvet, N., Gullbring, E., \& D'Alessio, P. 1998, \apj, 495, 385
\bibitem[Ho et al.(2004)]{ho04} Ho, P. T. P., Moran, J. M., \& Lo, K. Y. 2004, \apj, 616, L1
\bibitem[Hubickyj et al.(2005)]{hubickyj05} Hubickyj, O., Bodenheimer, P., \& Lissauer, J. J. 2005, Icarus, 179, 415
\bibitem[Hughes et al.(2007)]{hughes07} Hughes, A. M., Wilner, D. J., Calvet, N., D'Alessio, P., Claussen, M. J., \& Hogerheijde, M. R. 2007, \apj, 664, 536
\bibitem[Lada et al.(2006)]{lada06} Lada, C. J., et al. 2006, \aj, 131, 1574
\bibitem[Lay et al.(1997)]{lay97} Lay, O. P., Carlstrom, J. E., \& Hills, R. E. 1997, \apj, 489, 917
\bibitem[Makarov(2007)]{makarov07} Makarov, V. V. 2007, \apj, 670, 1225
\bibitem[Muzerolle et al.(1998)]{muzerolle98} Muzerolle, J., Hartmann, L., \& Calvet, N. 1998, \aj, 116, 2965
\bibitem[Najita et al.(2007)]{najita07} Najita, J. R., Strom, S. E., \& Muzerolle, J. 2007, \mnras, 378, 369
\bibitem[Natta et al.(2006)]{natta06} Natta, A., Testi, L., \& Randich, S. 2006, \aap, 452, 245
\bibitem[Pi{\'{e}}tu et al.(2006)]{pietu06} Pi{\'{e}}tu, V., Dutrey, A., Guilloteau, S., Chappilon, E., \& Pety, J. 2006, \aap, 460, L43
\bibitem[Ridge et al.(2006)]{ridge06} Ridge, N. A., et al. 2006, \aj, 131, 2921
\bibitem[Rodmann et al.(2006)]{rodmann06} Rodmann, J., Henning, T., Chandler, C. J., Mundy, L. G., \& Wilner, D. J. 2006, \aap, 446, 211
\bibitem[Stapelfeldt et al.(2008)]{stapelfeldt08} Stapelfeldt, K., et al. 2008, in preparation
\bibitem[Testi et al.(2001)]{testi01} Testi, L., Natta, A., Shepherd, D. S., \& Wilner, D. J. 2001, \apj, 554, 1087
\bibitem[Testi et al.(2003)]{testi03} --------- 2003, \aap, 403, 323
\bibitem[Vrba et al.(1993)]{vrba93} Vrba, F. J., Coyne, G. V., \& Tapia, S. 1993, \aj, 105, 1010
\bibitem[Weidenschilling(1977)]{weidenschilling77} Weidenschilling, S. J. 1977, \apss, 51, 153
\bibitem[Wilking et al.(2005)]{wilking05} Wilking, B. A., Meyer, M. R., Robinson, J. G., \& Greene, T. P. 2005, \aj, 130, 1733
\bibitem[Wilner et al.(1996)]{wilner96} Wilner, D. J., Ho, P. T. P., \& Rodgriguez, L. F. 1996, \apj, 470, L117
\bibitem[Wilner \& Lay(2000)]{wilner00} Wilner, D. J., \& Lay, O. P. 2000, in Protostars \& Planets IV, eds. V. Mannings, A. P. Boss, \& S. S. Russell (Tucson: Univ. Arizona Press), 509
\bibitem[Wilner et al.(2005)]{wilner05} Wilner, D. J., D'Alessio, P., Calvet, N., Claussen, M. J., \& Hartmann, L. 2005, \apj, 626, L109
\end{thebibliography}
\end{document}